\documentclass[prl, twocolumn, floatfix]{revtex4} 
\usepackage{graphicx}
\usepackage{amsmath, amsfonts, amssymb, bm}
\usepackage{braket}
\usepackage{color}
\usepackage[utf8]{inputenc}
\usepackage{subfigure}

\begin{document}

\title{ Breit interaction overtaking Coulomb force at low energies: 
an unexpectedly efficient mechanism for ionization in slow collisions } 

\author{ A. Jacob, C. M\"uller and A. B. Voitkiv}
\affiliation{ Institut f\"ur Theoretische Physik I, 
Heinrich-Heine-Universit\"at D\"usseldorf, 
\\ Universit\"atsstr. 1, 40225 D\"usseldorf, Germany }
\date{\today}

\begin{abstract}  

It is generally assumed that 
ionization in slow collisions of light atomic particles, whose constituents (electrons and nuclei) move with velocities orders of magnitude smaller than the speed of light, is driven solely by the Coulomb force. Here we show, however, that the Breit interaction -- a relativistic correction to the Coulomb interaction between electrons -- can become the main actor when the colliding system couples resonantly to the quantum radiation field. Our results demonstrate that this ionization mechanism can be very efficient in various not too dense physical environments, including stellar plasmas and atomic beams  propagating in gases. 

\end{abstract} 

\maketitle 

Being electrically charged, electrons communicate with each other via electroweak interactions \cite{EWI}.  When electron velocities are much smaller than the speed of light, the primary interaction between them is the (instantaneous) Coulomb force. The leading relativistic correction to it 
is the (generalized) Breit interaction \cite{gbi}, which normally is important for high-energy processes only \cite{f1}, \cite{r1}-\cite{r3}. 

Electron-electron interactions are crucial for a vast amount of phenomena in nature. They are, for instance, responsible for processes involving energy transfer between bound states of atoms and molecules, in which an electronically excited atomic site transitions to a lower lying state, transferring the released energy to a partner particle which in turn becomes excited. Examples of such processes  
include the energy transfer in condensed-matter systems \cite{cms}-\cite{cms1}, quantum optical ensembles \cite{qoe}  and cold Rydberg gases \cite{crg}-\cite{crg1}, and the population inversion in a He-Ne laser 
\cite{He-Ne-las}. They moreover comprise chemical and biological systems such as  lattice dynamics in polymers \cite{ldp}-\cite{ldp1} and F\"orster resonances between chromophores \cite{chromophore}-\cite{chromophore1}. 

These interactions also drive autoionization occurring when an atomic (or molecular) system 
occupies an electronically excited (quasi-) bound state, whose excitation energy is larger than its ionization potential: the system then can relax by emitting an electron. Autoionization, in which -- in a general case -- a system 'spontaneously' decays into 
its (charged) constituent parts, has many different facets, playing an important role in various physical \cite{auto-phys},  
chemical \cite{auto-chem} and biological environments 
\cite{f2}.  

In a system, consisting of different atomic particles,   
electronic excitation in one 
of them can be transformed into ionization of another one. 
If the de-excitation transition is optically allowed, autoionization occurs also at large inter-particle distances, where it is driven by the long-range dipole-dipole interaction 
and proceeds via energy transfer between the particles. 

Such a two-center autoionization 
is very efficient in weakly bound systems and thermal collisions. For instance, the inter-atomic Auger decay \cite{iaad}, where a filling of a vacancy in an exited atom proceeds via 
ionization of a neighbor atom,  
can even outperform the 'normal' (intra-atomic) Auger decay. Also, 
collisional quenching of optically excited atoms at thermal velocities \cite{firsov} has much larger cross sections compared to 
those of metastable atoms (Penning ionization) 
and ionization by electron impact. 

Recent years witnessed a rapid growth in the studies of the various aspects of two-center autoionization, which showed its  importance for a multitude of physical, chemical and biological systems (for recent reviews see \cite{recent-icd-reviews}-\cite{recent-icd-reviews1}).   
This renewed interest in two-center autoionization was triggered by the results of \cite{icd} that treated   
the process of inter-atomic Auger decay in a system of two atomic particles, where relaxation via intra-atomic Auger decay is forbidden energetically, and introduced the name interatomic coulombic decay for it \cite{f3}.  

The basic feature of all inter-atomic autoionization processes  
(both in weakly bound systems and slow collisions), considered up to now, is that they are strongly  dominated by the Coulomb interaction whereas relativistic corrections to this interaction are not important 
\cite{we-2cpi-2010-paper, nat-comm, we-2021, TK}   
(unless highly charged ions are involved \cite{nat-comm}, where electrons move with very large velocities).   
This overwhelming dominance is nicely emphasized by 
the term 'interatomic {\it coulombic} decay'.  
 
In this communication we introduce a hitherto unexplored mechanism of autoionization in very slow collisions of atomic particles. It is driven by a particular form of the Breit interaction, arising when the colliding system couples resonantly to the quantum radiation field, and proceeds via energy transport between the particles, having an extremely long range. This mechanism turns out to be so surprisingly efficient that it can outperform all coulombic mechanisms of ionization in various not too dense environments.  

Atomic units ($\hbar = |e| = m_e = 1$) 
are used throughout unless otherwise stated.  

\vspace{0.25cm}  

Let a beam of atomic species A, moving with a low velocity $v$ ($v \ll 1$ a.u. $ \approx 2.18 \times 10^{8}$ cm/s), penetrate a gas of atoms B. Suppose that initially 
atoms B are in an excited internal state $ u_1 $ 
with an energy $ \epsilon_1 $ and that the excitation energy 
$\omega_{\rm B} = \epsilon_1 - \epsilon_0 $, where 
$\epsilon_0$ is the energy of the ground state $u_0$ of B, is larger than the ionization potential $ - \varepsilon_0 $ of A. 
If $u_1 \leftrightarrow u_0 $ is an electric dipole transition,  
then an efficient pathway for ionization of A  
in the collision is to get the energy $\omega_{\rm B} $ stored in atom B (see Fig. 1 for illustration). 

\begin{figure}[h!]
\centering
\includegraphics[width=7.cm]{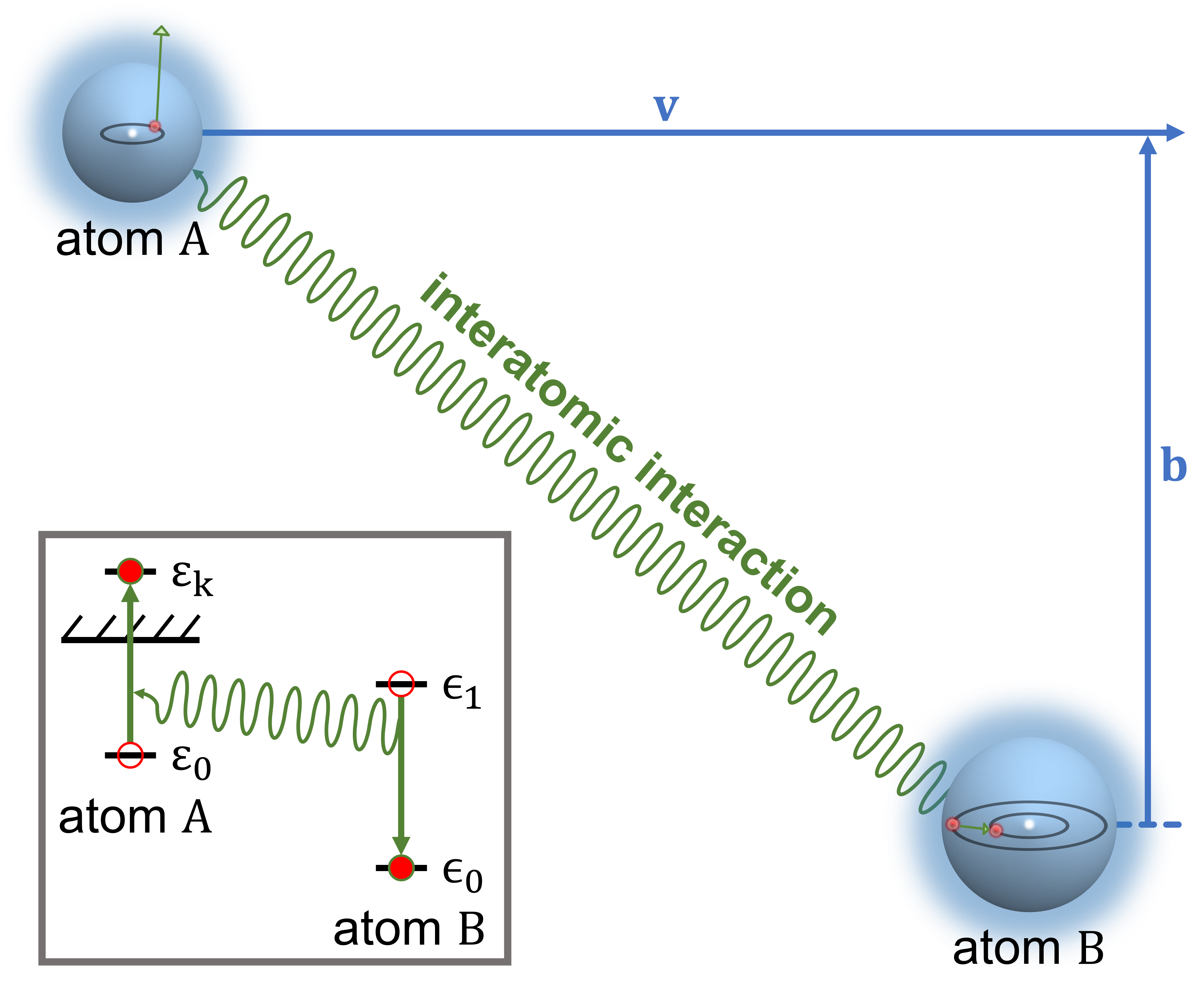}
\caption{ Scheme of ionization in collisions with excited atoms. }
\end{figure}

We assume that $ v \ll v_e \ll c$, where $v_e \sim 1$ a.u. are the characteristic velocities of the 'active' electrons of A and B and $ c \approx 137 $ a.u. is the speed of light and give a QED description of the collisions between A and B. According to QED the (instantaneous) Coulomb interaction arises due to the exchange of time-like and longitudinal photons  
whereas the Breit interaction is caused by the exchange of transverse photons. These two interactions are characterized by  different selection rules and the total cross section $\sigma_{\rm{t}}$ for ionization of A reads     
\begin{eqnarray} 
\sigma_{\rm t} = \sigma_{\rm Coul} + 
\sigma_{\rm Br},    
\label{v5-total} 
\end{eqnarray} 
where $\sigma_{\rm Coul}$ and $\sigma_{\rm Br}$ are the contributions to the total cross section due to 
the Coulomb and Breit interactions, respectively. 
The term $\sigma_{\rm Br}$ arises because -- in contrast 
to  the 'standard' theories (e.g. \cite{mcdowell,sdr,mcguire}) of slow  
collisions of light atomic particles --    
we do not approximate the electromagnetic field transmitting the interaction between A and B 
by an instantaneous Coulomb form 
but describe it relativistically.

Compared to the Coulomb force   
the Breit interaction contains an extra factor $\sim ( v_e/c)^2 $, which in our case is very small. Therefore, this interaction can become a significant actor only provided the smallness of this factor can be compensated by the resonant character of 
the coupling between  
the colliding A-B system and the radiation field. 
The latter is the case for ionizing collisions, which populate 
the interval of electron emission energies 
$\varepsilon_k$ given by 
\begin{eqnarray} 
\varepsilon_0 + \omega_{\rm B} \left(1 - \frac{v}{c} \right)  \leq   \varepsilon_k \leq \varepsilon_0 + \omega_{\rm B} \left(1 + \frac{v}{c} \right),     
\label{v5-res-en} 
\end{eqnarray}
where the Breit interaction acquires an extremely long 
range being transmitted by on-shell photons.  
Thus, the Breit interaction 
can only be 'visible' in that part of ionization of A which proceeds with electron emission into the energy interval (\ref{v5-res-en}). 
On the other hand, since the interval (\ref{v5-res-en}) is very narrow, it contributes negligibly to the ionization caused by the Coulomb interaction.   

\vspace{0.25cm} 

{\bf { A. Ionization via the Breit interaction.}}  
As was mentioned,  
the population of the interval of electron emission energies given by Eq. (\ref{v5-res-en}) 
proceeds via the exchange of a photon, whose  four-momentum $ q^\mu $ satisfies the on-shell condition  
$ q_\mu q^\mu = 0 $. In the rest frame of B its frequency is $\omega_{\rm B}$ but in the rest frame of A its frequency 
$\omega'_{\rm B}$ is Doppler shifted occupying the range $ \vert \omega'_B - \omega_{\rm B} \vert \leq \omega_{\rm B} \, v/c$. 

When $v \ll c$, one can derive a simple formula for 
the contribution $ \sigma_{\rm Br} $ of the Breit interaction to the total cross section for ionization of A   
\begin{eqnarray}     
\sigma_{\rm Br} & = &  \alpha_{\rm Br} \, \,   
\frac{ \Gamma_{\rm r}^{\rm B} \, 
\sigma_{\rm ph}^{\rm A}(\omega_{\rm B}) \, b_{\rm max}}{ v }.    
\label{v5-breit} 
\end{eqnarray} 
Here, $b_{\rm max}$ ($b_{\rm max} \gg c/\omega_{\rm B}$) is the maximum value of the impact parameter, 
$ \Gamma_{\rm r}^{\rm B}$ is the natural width of the excited state of B and 
$\sigma_{\rm ph}^{\rm A}(\omega_{\rm B})$ is the cross section for ionization of atom A by absorption of a photon with energy $ \omega_{\rm B}$ \cite{f4}.  
$\alpha_{\rm Br}$ is a numerical factor  dependent on the electron transition in B: for instance, quantizing the states along the collision velocity, 
one obtains $\alpha_{\rm Br} = 3 \pi/8 $ and $\alpha_{\rm Br} = 9 \pi/ 16 $ for $2p_0 \to 1s$ and $2p_{\pm 1} \to 1s$, respectively.      

The dependence   
$ \sigma_{\rm Br} \sim \Gamma_{\rm r}^{\rm B} \, 
\sigma_{\rm ph}^{\rm A}  
\, b_{\rm max} / v $ arises due to a peculiar origin of the Breit interaction in the case under consideration where it is caused by the {\it resonant} coupling of the colliding A--B system to the radiation field (that is also responsible for spontaneous radiative decay of the excited state of B) and the contribution of this mechanism  
to the ionization rate at a fixed distance $R$ between A and B, when $R \gg c/\omega_{\rm B} $,  is proportional to $\Gamma_{\rm r}^{\rm B} \sigma_{\rm ph}^{\rm A}/R^2$.  
Accordingly, this Breit interaction 
and the corresponding ionization mechanism can be termed 'resonant Breit interaction' and 
'resonant Breit ionization', respectively.  
 
Since the overwhelming contribution to the cross section $\sigma_{\rm Br}$ arises from collisions with extremely large impact parameters, 
Eq. (\ref{v5-breit}) remains valid also at very low impact energies ($ v \to 0 $) with the corresponding rate constant 
$v \, \sigma_{\rm Br}$ being independent of the energy. 

\vspace{0.25cm} 

{\bf { B. Ionization via the Coulomb interaction.} }
In distant collisions, 
where the electronic shells of A and B always remain very well separated in space and the Coulomb interaction between the atoms is weak, the latter is transferred  
by a single off-shell photon.  
In such collisions, the corresponding (first order) Coulomb  ionization probability 
$\mathcal{P}_{\rm Coul}(b)$ at $v \ll 1$ reads 
\begin{eqnarray} 
\mathcal{P}_{\rm Coul}(b) & = &  \frac{ 3 }{ 2 \pi } \, \alpha_{\rm Coul}^{} \, \left( \frac{ c }{ \omega_{\rm B}} \right)^4 \, \frac{ \Gamma_{\rm r}^{\rm B} \, \sigma_{\rm ph}^{\rm A}(\omega_{\rm B}) }{ v \, \, b^5 }.         
\label{v5-coul-prob}
\end{eqnarray}
The corresponding contribution $\sigma^{\rm dist}_{\rm Coul}$ to the Coulomb cross section $\sigma_C$ is given by 
\begin{eqnarray}
\sigma^{\rm dist}_{\rm Coul} 
& = &  \alpha_{\rm Coul}^{} 
\left(\frac{ c }{ \omega_{\rm B} } \right)^4 
\, \frac{ \Gamma_{\rm r}^{\rm B} \, 
\sigma_{\rm ph}^{\rm A}(\omega_{\rm B}) }{ v \, b_0^3 }.      
\label{v5-coul}
\end{eqnarray}
Here, $\alpha_{\rm Coul} = 9 \pi/64 $ ($2 p_0 \to 1s$) 
or $\alpha_{\rm Coul} = 27 \pi/128 $ ($2 p_{\pm 1} \to 1s$) and $ b_0 $ is some minimum value of the impact parameter. The form of 
$ \mathcal{P}_{\rm Coul}(b) $ and $\sigma^{\rm dist}_{\rm Coul}$ points to the fact that in this case we deal with 
the 'coulombic' two-center autoionization (interatomic coulombic decay)  
whose rate $\Gamma_a $ at large inter-atomic distances ($R \gg 1$ a.u.) is proportional to  
$ (c/\omega_{\rm B})^4 \, \Gamma_{\rm r}^{\rm B} \, \sigma_{\rm ph}^A(\omega_{\rm B}) /R^6  $ \cite{iaad}.   

If the magnitude of $\mathcal{P}_{\rm Coul}(b) $ approaches, or even exceeds, $1$ at $b \gg 1$ a.u. (i.e. the first order approximation breaks down even at large $b$), then one can obtain a simple estimate for the total Coulomb cross section $ \sigma_{\rm Coul} $    
\begin{eqnarray} 
\sigma_{\rm Coul}  
& = & 3.48 \, \, \alpha_{\rm Coul}^{2/5}  \, \left( \frac{ c }{ \omega_{\rm B}} \right)^{1.6} \, \,   
\bigg[\frac{ \Gamma_{\rm r}^{\rm B} \, 
\sigma_{\rm ph}^{\rm A}(\omega_{\rm B}) }{ v }\bigg]^{2/5}.      
\label{v5a-coul}
\end{eqnarray} 
Eq.(\ref{v5a-coul}) can be used if the cross section values, that it provides, are much larger than 
$ \pi a_0^2$, where $a_0 = 1$ a.u. 
($ \pi a_0^2  \simeq 10^{-16}$ cm$^2$). 

The cross section estimates of the form, given by Eq.  (\ref{v5a-coul}), yield good results at very low impact velocities $v$ \cite{firsov}. With increasing $v$   
(but still $v \ll 1$ a.u.) eventually the cross section (\ref{v5-coul}) with an appropriately chosen $b_0$ 
is expected to become a better approximation for 
$ \sigma_{\rm Coul} $.  
 
\vspace{0.25cm} 

{\bf { C. Ionization cross sections and rates}.}
\begin{figure}[h!]
\centering
\subfigure{\includegraphics[width=4.3cm]{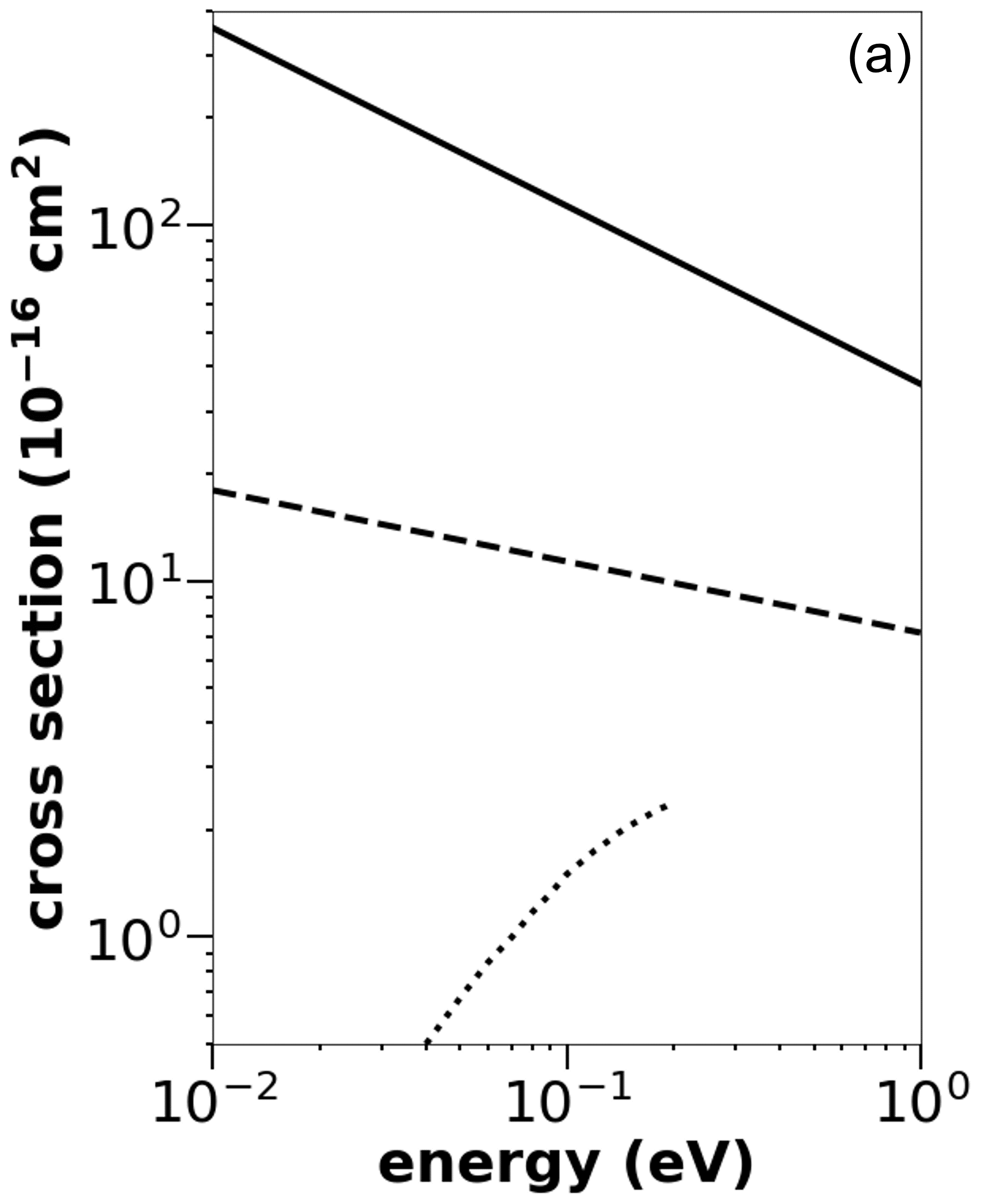}}
\subfigure{\includegraphics[width=4.1cm]{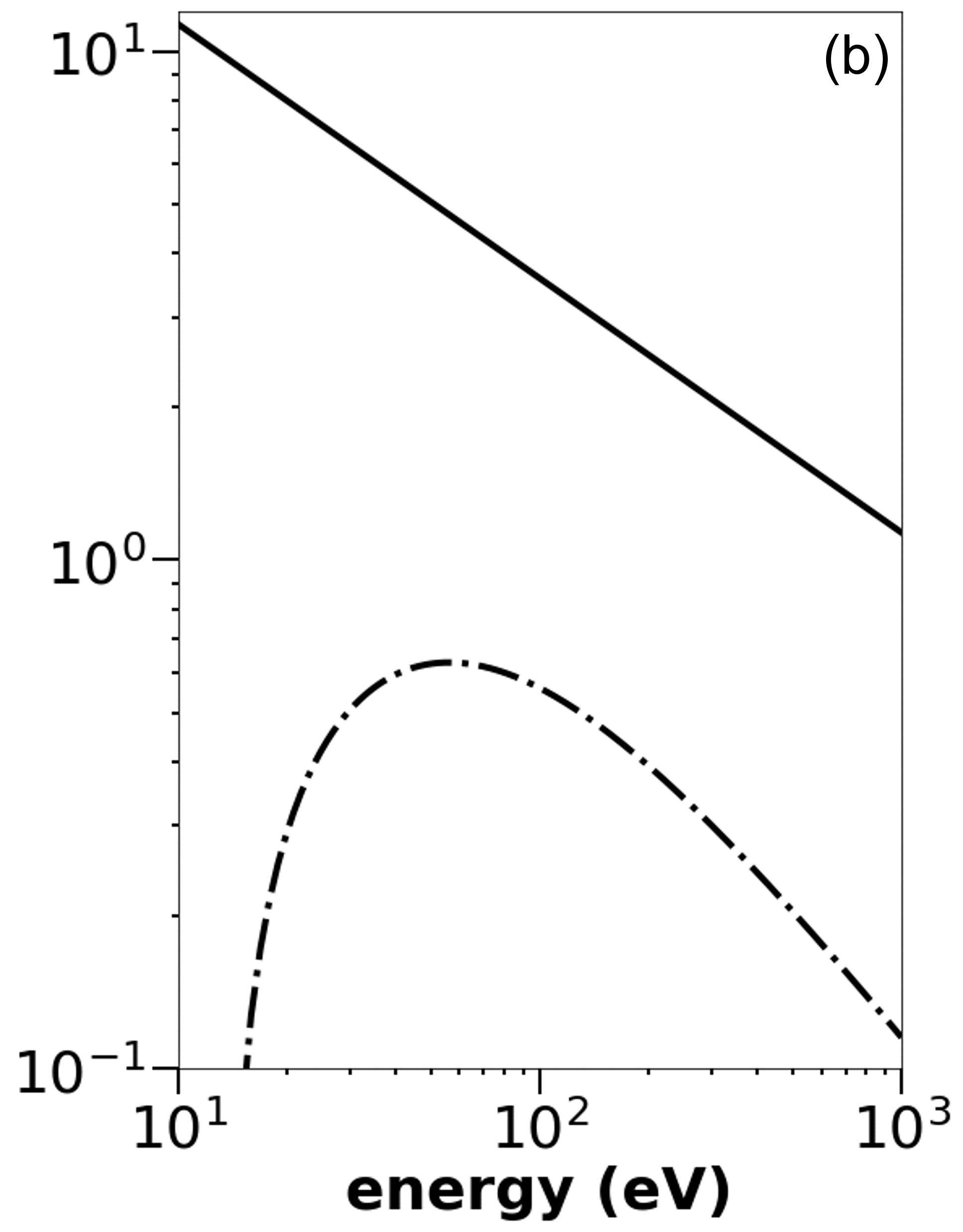}}
\caption{ The cross sections for ionization of H(1s) by 
He(2\,$^1$P), He(2\,$^1$S) and free electrons given as a function of the relative kinetic energy. 
Solid curves in (a) and (b): He(2\,$^1$P), 
the cross section $\sigma_{\rm Br}$ 
(Eq.(\ref{v5-breit}) with $b_{\rm max} = 1$ cm). 
Dash curve in (a): He(2\,$^1$P), 
the cross section $\sigma_{\rm{Coul}}$ (Eq.(\ref{v5a-coul})).   
Dot curve in (a): experimental data for Penning ionization by He(2\,$^1$S) (from \cite{penn}). 
Dash-dot curve in (b): ionization by 
free electrons. 
The results for collisions with He(2$^1$P)
are averaged over the magnetic quantum number of the excited state. }
\label{figure2}
\end{figure}
Let us now apply our theory to collisions involving hydrogen and helium, which are the most abundant elements in the universe \cite{f5}. 
In low-energy collisions of ground-state atoms  
ionization is not likely. 
However, collisions with free electrons and Penning ionization can be important sources of ionization in gases and low-temperature plasmas. 
Therefore, in Fig. 2, in addition to the ionization cross sections in collisions of H(1s) (atom A) with He(2$^1$P),  we show also results for ionization of H(1s) by free electrons \cite{f6}, \cite{el-imp-1}   
and by Penning ionization 
in collisions with He(2$^1$S).  

It is seen that the cross section for ionization in collisions with He(2$^1$P), dominated by the Breit interaction, is much larger than those for Penning ionization and ionization by electron impact.  

The results for $\sigma_{\rm Br}$, shown in Fig. 2, 
were obtained by choosing $b_{\rm max} = 1$ cm. In general, 
one has $b_{\rm max} \leq \Lambda$, where $\Lambda$ is the effective range of the Breit interaction. In collisions in a not too dense H--He gas, containing a fraction of excited 
He atoms, $\Lambda $ is determined by the mean free path of radiation at photon energy of $\approx 21.2$ eV (resonant to the $1 \, ^1 S \to 2\, ^1 P$ transition in He). 
Our estimates show, for example, that 
at $T = 300$ K, $T = 11600$ K ($\approx 1$ eV) 
and $T = 20000$ K the magnitude of $ \Lambda $ exceeds $1$ cm at 
$n_B \lesssim 10^{15}$ cm$^{-3}$, 
$ n_B \lesssim 4 \!\times\! 10^{16} $ cm$^{-3}$ 
and $n_B \lesssim 7 \!\times \! 10^{16} $ cm$^{-3}$, respectively.  

Considering the reaction 
H(2s) + H(2p) $\to $ H$^+$ + H(1s) + e$^-$   
and choosing $b_{\rm max} = 1$ cm we get $\sigma_{\rm Br} = 6.14 \!\times\! 10^{-15}$ cm$^2$ 
($\sigma_{\rm Coul} = 3.5 \!\times\! 10^{-15}$ cm$^2$) and 
$\sigma_{\rm Br} = 8.7 \!\times\! 10^{-16}$ cm$^2$ ($\sigma_{\rm Coul} = 1.6 \!\times\! 10^{-15}$ cm$^2$) for the relative kinetic energies of $0.01 $ and $0.5 $ eV, respectively \cite{f7}.  

Further, for neutralization of H$^-$ in collisions with H(2p), H$^-$ + H(2p) $\to $ 2 H(1s) + e$^-$,  
for the same relative kinetic energies, $0.01 $ and $0.5 $ eV, and setting again $b_{\rm max} = 1$ cm we obtain $\sigma_{\rm Br} = 2.5 
\!\times\! 10^{-14}$ cm$^2$ 
($\sigma_{\rm Coul} = 6.2 \!\times\! 10^{-15}$ cm$^2$) and 
$\sigma_{\rm Br} = 3.6 \!\times\! 10^{-15}$ cm$^2$ ($\sigma_{\rm Coul} = 2.8 \!\times\! 10^{-15}$ cm$^2$), respectively \cite{f8}, \cite{Hmin}.  

Yet another similar reaction is ionization of He(1s$^2$) in collisions with He$^+$(2p): He(1s$^2$) + He$^+$(2p) $\to $ 
2 He$^+$(1s) + e$^-$. 
For the relative kinetic energies of $0.01 $ and $0.5 $ eV 
(and $b_{\rm max} = 1$ cm) we obtain 
$\sigma_{\rm Br} = 4.74 \!\times\! 10^{-13}$ cm$^2$ 
($\sigma_{\rm Coul} = 21.6 \!\times\! 10^{-16}$ cm$^2$) and 
$\sigma_{\rm Br} = 6.7 \!\times\! 10^{-14}$ cm$^2$ 
($\sigma_{\rm Coul} = 9.9 \!\times\! 10^{-16}$ cm$^2$), respectively \cite{f9}, \cite{Marr}. 

\vspace{0.25cm} 
The above results demonstrate that the resonant Breit interaction can lead to surprisingly high cross section values. Let us now consider whether this interaction 
can also result in substantial reaction rates in 
such physical environments as (stellar) plasmas and beams of particles injected into gases.  

As the first example we briefly 
consider an electrically neutral hydrogen--helium plasma with the densities $n_{\rm H} = 3 \times 10^{15}$ cm$^{-3}$ 
($5 \times 10^{-6}$ kg/m$^3$) 
and $n_{\rm He} = \frac{1}{10} \times n_{\rm H} = 3 \times 10^{14}$ cm$^{-3}$ ($ 2 \times 10^{-6}$ kg/m$^3$), respectively, which has a temperature $T = 14500$ K. Using the Maxwell and Boltzmann distributions and the Saha equation we obtain that the plasma contains neutral hydrogen (HI) and protons (HII)
in a proportion $n_{\rm HI}/n_{\rm HII} \approx 1/21$ 
and mostly neutral helium ($\approx 98.3 \% $ of helium, predominantly in the ground state) with a tiny fraction ($\simeq 10^{-7}$) of He(2$^1$P) (the corresponding density is $ \approx 3.8 \times 10^{7}$ cm$^{-3}$). Setting $b_{\rm max} = \Lambda$, where $\Lambda \approx 179$ cm is the calculated effective range of the Breit interaction \cite{f10}, 
we obtain that the rate for ionization of H(1s) via this interaction is $\approx 38$ s$^{-1}$. This can be compared to the rate 
of $\approx 346$ s$^{-1}$ for ionization of H(1s) by plasma electrons and the rate of $\approx 370$ s$^{-1}$ for radiative electron-proton recombination into H(1s). We thus see that the resonant Breit interaction results in quite a 'visible' ionization rate.     

\vspace{0.15cm} 

As the second example let us explore ionization 
of a beam of atoms A moving in a gas of atoms B  
which are excited by a weak laser field of frequency $\omega$ nearly resonant to transitions in atoms B ($\omega \approx \omega_{\rm B}$).  
We shall term this process collisional two-center photo ionization (collisional 2CPI or simply 2CPI), following \cite{we-2cpi-2010} where a similar ionization process was predicted for a system of two weakly bound atomic species and its very high efficiency was shown, that was confirmed in experiments on photo ionization of Ne-He dimers \cite{fr-exper} and Ar-Ne clusters \cite{Ar-Ne-clusters}.   

The cross sections for 2CPI can be obtained by multiplying the cross sections (\ref{v5-breit}) and (\ref{v5-coul})-(\ref{v5a-coul}) by the probability $p$ for finding atoms B in the excited state \cite{f11}.  
Assuming that the steady-state condition, 
$dp/dt = 0$, 
is fulfilled for atoms B interacting 
with the laser field and, accordingly, 
one has $ \Gamma_{\rm r}^{\rm B} \, p = n_{\rm ph} \, \sigma_{\rm sc}^{\rm B} \, c$, where 
$\sigma_{\rm sc}^{\rm B}$ is the cross section for resonant scattering of laser photons on B and 
$n_{\rm ph} $ is the photon density, we obtain    
for the contribution 
$\sigma_{\rm 2CPI}^{\rm Br}$ to the cross section for 2CPI due to the Breit interaction   
\begin{eqnarray} 
\sigma_{\rm 2CPI}^{\rm Br} = \alpha_{\rm Br} \, \, 
\bigg[ n_{\rm ph} \, c \, \frac{b_{\rm max}}{v} \, \sigma_{\rm sc}^{\rm B} \bigg] \, \sigma_{\rm ph}^{\rm A}(\omega).      
\label{v7-2cpi-breit} 
\end{eqnarray}

The process of 2CPI competes with the direct photo ionization (DPI) of atoms A by the laser field. The relative effectiveness of these two processes can be described by the ratio of the (total) rate $ \mathcal{R}_{\rm 2CPI} = n_B \, v \, 
\sigma_{\rm 2CPI} $ for 2CPI 
($n_{\rm B}$ is the density of atoms B)  
and the rate $\mathcal{R}^{\rm A}_{ \rm DPI} = n_{\rm ph} \, c \, \sigma_{\rm ph}^{\rm A}(\omega)$ for direct photo ionization. 
Assuming that 2CPI is dominated by the Breit interaction we obtain 
\begin{eqnarray} 
\zeta & = & \frac{ \mathcal{R}_{\rm{2CPI}}} { 
\mathcal{R}^{\rm A}_{ {\rm DPI} } }  
\approx \alpha_{\rm Br} \, \, \frac{ b_{\rm max} }{ \Lambda_{\rm rad}^{\rm B} },    
\label{v8} 
\end{eqnarray} 
where $ \Lambda_{\rm rad}^{\rm B} = 1/(n_{\rm B} \,  \sigma_{\rm sc}^{\rm B}) $ 
is the mean free path of the radiation in the gas of atoms B.  

According to Eq. (\ref{v8}), $\zeta$ is determined by   
the ratio of two distances, $b_{\rm max}$ and $\Lambda_{\rm rad}^{\rm B} $.  The magnitude of $b_{\rm max}$ depends on the size of the target gas of atoms B and/or the size of the beam of atoms A. However, its maximum value obviously cannot considerably exceed $\Lambda_{\rm rad}^{\rm B}$ and, in the case $b_{\rm max} = \Lambda_{\rm rad}^{\rm B}$, we obtain 
$ \zeta = 3 \pi/8 $ (or $ \zeta = 9 \pi/16 $). 
This means that the resonant Breit interaction can have a significant effect on the ionization rate for atoms A.  

\vspace{0.15cm} 
Summarizing the results for the cross sections and rates 
presented in this subsection 
we can draw a conclusion about a surprisingly high ionization efficiency of the resonant Breit interaction. 
This is not only interesting from the point of view of basic physics but also can make this interaction an important factor in various not too dense physical environments, including atomic/molecular beams propagating in  
gases and astrophysical objects like stellar plasmas, which consist primarily of hydrogen and helium and for which such parameters as the temperature and particle densities vary in very broad ranges \cite{aph1}-\cite{aph3}. 

We also note that the effect of the Breit interaction in collisional 2CPI can be explored experimentally by  
using a beam of slow atoms/ions (e.g. H, H$^-$, H$_2^+$, Mg$^+$, Ca$^+$, etc.)  
and a gas (e.g. He, Ne) target irradiated 
by a weak (resonant) laser field. 
Moreover, one can envisage experiments in which  
the projectile beam passes close by the target without penetrating it and being not exposed to the laser field. Then the Coulomb interaction between A and B as well as the interaction of A with the laser field are excluded and only the resonant Breit interaction can contribute to ionization of A. 

\vspace{0.25cm} 
{\bf {D. Resonant Breit ionization  
and the process of resonance energy transfer.} }
Resonance energy transfer is a process, where the excess energy of an excited molecule (atom) is transferred to another molecule (atom) which gets excited (for a recent review on this topic see \cite{RET}).  
It is known that in order to properly describe this process, when it occurs at very large intermolecular distances, one has to take into account the coupling of the molecules to the  radiation field. In this respect the resonance energy transfer is similar to the resonant Breit ionization in slow collisions, where such coupling also plays the crucial role. Thus, our present results point to an interesting connection between these two research fields. 
 
\vspace{0.25cm} 

{\bf {E. Resonant Breit ionization versus  
resonant high-energy processes. } }
Ionization in slow collisions is a low-energy process where both the collision energy and the energy transfer between the atoms are very far from the corresponding relativistic domains. It is, therefore, of interest to note that its mechanism  has important similarity to the cascade mechanism of $e^+ e^-$ pair production \cite{slac}-\cite{slac1}, \cite{pp-rev}  
and the resonant mechanism of 
the excitation of atoms \cite{we-2009} by high-energy electrons moving in a strong laser field. The reason is that all these mechanisms become possible due to the resonant coupling to the radiation field.  
 
\vspace{0.25cm} 

{\bf { Conclusions. } }
Ionization of atoms A in slow collisions with excited atoms B, where the energy stored in B before the collision exceeds the ionization potential of A and is released via a dipole allowed transition, can be strongly dominated  
by the Breit interaction.  

This is quite spectacular since this interaction, being a relativistic correction to the Coulomb one, has so far been regarded as fully irrelevant for the field of low-energy collisions. 
However, in collisions with excited atoms  
when the condition $ \vert \varepsilon_k - \varepsilon_0 - \omega_{\rm B} \vert \leq v \omega_{\rm B}/c $ is fulfilled, the A--B system couples resonantly to the quantum radiation field. This tremendously increases the range of the Breit interaction, strongly enhancing its efficiency. 

The latter was demonstrated by calculating ionization cross sections and rates, showing that the Breit interaction 
can play an important role in various not too dense physical environments. 
 
\begin{acknowledgements}  
We acknowledge useful discussions with Prof. K.-H. Spatschek on astrophysical plasmas.
\end{acknowledgements}


\begin{thebibliography}{33}
\bibitem{f1}
The (main) correction to the atomic energy levels due to the Breit interaction   
contains an additional factor of $(Z \alpha)^2$, 
where $\alpha \approx 1/137$ is the fine structure constant,  
and for light atomic systems ($ Z \sim 1$) it is roughly by $4$ orders of magnitude smaller than the contribution 
from the Coulomb interaction between the electrons. In low-energy collisions of light atomic species the Breit interaction was, to our knowledge, never considered, being tacitly assumed to be of negligible importance. 
In the case of high-energy (relativistic) electrons the significance of the Breit  interaction naturally greatly increases, both for bound systems and collisions. In particular, the role of the Breit interaction in collision processes involving very highly charged ions has attracted much attention \cite{r1}-\cite{r3}.
\bibitem{f2} For instance, the autoionization (dissociation) of water, H$_2$O $ \to $ HO$^-$ + H$^+$, 
a fundamental chemical reaction occurring in the aqueous phase, is of major importance in many areas of chemistry and biology.
\bibitem{f3} The term 'interatomic coulombic decay' is now often understood more generally and used to denote all long-range interatomic/intermolecular autoionization mechanisms, 
which at large distances are driven by the dipole-dipole interaction.
\bibitem{f4} One should note that the quantity $ \sigma_{\rm Br} $ depends on the collision geometry 
($b_{\rm max}$) and, therefore, is not a 'normal' cross section.
\bibitem{f6} The cross section for ionization of H(1s) by electron impact was calculated using results of \cite{el-imp-1}.
\bibitem{f7} We neglected 
the channel of Penning ionization 
which is expected to be weak.
\bibitem{f8} The value of $\sigma_{\rm ph}^{\rm A}$ for photo detachment from H$^-$ was taken from \cite{Hmin}.
\bibitem{f9} The value of $\sigma_{\rm ph}^{\rm A}$ for photo ionization of He(1s$^2$) by a photon with $40.8$ eV energy was taken from \cite{Marr}.
\bibitem{f10} In the case under consideration 
$\Lambda$ is determined mainly by the (resonant) interaction of the radiation at $\approx 21.2$ eV with He(1 $^1$S) atoms.
\bibitem{f11} A more detailed consideration shows that 
in the cross sections (\ref{v5-breit}), 
(\ref{v5-coul})-(\ref{v5a-coul}) the quantity $\omega_{\rm B}$ should be replaced by the frequency of the laser field $\omega$.
\end{thebibliography}
\end{document}